\documentclass[aps,prb,twocolumn,showpacs]{revtex4}
\usepackage{epsfig}
\usepackage{hyperref}

\begin{document}

\def\il{I_{low}} 
\def\iu{I_{up}} 
\def\eeq{\end{equation}}
\def\ie{i.e.}  
\def\etal{{\it et al. }}  
\def\prb{Phys. Rev. {\bf B}}
\def\pra{Phys. Rev. {\bf A}} 
\def\prl{Phys. Rev. Lett. }
\def\pla{Phys. Lett. A } 
\def\pb{Physica B}
\def\ajp{Am. J. Phys. } 
\def\jpc{J. Phys. C } 
\def\rmp{Rev. of Mod. Phys. } 
\def\jap{J. Appl. Phys. } 
\def\mpl{Mod. Phys. Lett. {\bf B}} 
\def\ijmp{Int. J. Mod. Phys. {\bf B}} 
\def\ijp{Ind. J. Phys. }
\def\ijpap{Ind. J. Pure Appl. Phys. }
\def\ibmjrd{IBM J. Res. Dev. }
\def\pjp{Pramana J. Phys.}

\title{Dephasing via stochastic absorption: A case study in Aharonov-Bohm oscill  ations}
\author{Colin Benjamin}
\email{colin@iopb.res.in}
\author{A. M. Jayannavar}
\email{jayan@iopb.res.in}
\affiliation{Institute of Physics, Sachivalaya Marg, Bhubaneswar 751 005,
  Orissa, India}
\date{\today}

\begin{abstract}
   The Aharonov-Bohm ring has been the mainstay of mesoscopic
   physics research since its inception.  In this paper we have dwelt
   on the problem of dephasing of AB oscillations using a
   phenomenological model based on stochastic absorption. To calculate
   the conductance in the presence of inelastic scattering we have
   used the method due to Brouwer and Beenakker. We have shown that
   conductance is symmetric under flux reversal and visibility of AB
   oscillations decay to zero as a function of the incoherence
   parameter thus signalling dephasing in the system.  Some comments
   are made on the relative merits of stochastic absorption with
   respect to optical potential model, which have been used to mimic
   dephasing.
\end{abstract}


\maketitle

Dephasing is defined as the process by which quantum mechanical
interference is gradually destroyed\cite{imry,fluorian}. Dephasing of
electronic phase coherence may be caused by it's interaction with
other quasi particles in the system. The notion of intrinsic
decoherence and dephasing of a particle interacting with it's
environment is being investigated intensively. For, a comparison
between theory and experiment, it is necessary to know how dephasing
affects the various quantum phenomena related to transport.  In our
present work we will be interested in modeling dephasing
phenomenologically. The Aharonov-Bohm (AB) interferometer is one of
the best examples for analyzing how quantum interference effects are
affected by dephasing. In this interferometer the phase of the
electrons passing through the arms is modulated by the magnetic flux.
It is not unlike the Young's double slit experiment apart from the
presence of the magnetic flux. In the Young's double slit
interferometer as we know the intensity is given by
$I=|\Psi|^2=|\psi_1|^2+|\psi_2|^2+2 Re(\psi_1^*\psi_2^{} e^{i \phi})$,
the part $2 Re(\psi_1^*\psi_2^{} e^{i \phi})$ represents the
interference term. Here $\psi_1$ and $\psi_2$ are the complex wave
amplitudes across the upper and lower arms of the interferometer (this
discussion is restricted to only two partial amplitudes for
simplicity) and $\phi$ is phase difference between the two wave
amplitudes.  If there is no phase relationship between the waves then
the intensity will be $I=|\psi_1|^2+|\psi_2|^2$. Dephasing thus
suppresses the interference terms. In other words dephasing, leads to
the vanishing of the off-diagonal elements of the density matrix.
  
Aharonov-Bohm effect in mesoscopic rings manifests itself as a
periodic behavior of conductance as a function of magnetic flux
piercing the loop\cite{imry}. In an AB interferometer the electrons
are treated quantum mechanically along the ring and they retain their
phase memory across the entire sample and this gives rise to an
operative definition of the phase coherence length $L_{\phi}$ or
incoherence length.  On increasing the temperature $L_\phi$ becomes
less than the size of the ring and the phase coherence in the system
gradually disappears and as a consequence the AB effect (the
visibility of the AB oscillations) vanishes. Typically $L_{\phi}$
scales with temperature $T$ in a power law form, i.e.,
$L_{\phi}=T^{-p}$ ($p$ lies in the range $1$ to $2$).

  There are different ways to model dephasing in such mesoscopic
  systems. A controlled way to introduce dephasing in the AB
  interferometer is to attach a voltage probe\cite{buti} to the sample
  as in inset of Fig.~1 (Buttiker's model). This voltage probe
  breaks the phase coherence by removing electrons from the phase
  coherent channel in the system and subsequently re-injecting them
  without any phase relationship. Another way to introduce dephasing,
  is to add a spatially uniform imaginary potential ($-i V_i$) (or
  optical potential) to the Hamiltonian\cite{yohta}, which removes
  particles from the phase coherent motion in the system. In this case
  absorbed particles are identified as a spectral weight lost in the
  inelastic channels and they are re-injected back into the sample. In
  the method due to Zohta and Ezawa\cite{yohta} the total transmission
  is defined after re-injection as the sum of two contributions one
  due to the coherent part and the other due to the incoherent part,
  i.e., $T_{tot}=T_{coh}+T_{incoh}$. The incoherent part is calculated
  as $T_{incoh}=\frac{T_{r}}{T_{l}+T_{r}}A$, herein $T_{r}$ and
  $T_{l}$ are the probabilities for right and left transmission from
  the region of inelastic scattering and $A$ is the absorbed part
  which is given by $A= 1-T_{coh}-R_{coh}$. This model has been used
  by several other authors as well to simulate inelastic scattering.
  But, this model has a problem, in the presence of magnetic flux it
  is shown to violate the Onsager's two terminal symmetry
  relations\cite{pareek}, i.e., $T_{tot}(\Phi)\neq T_{tot}(-\Phi)$,
  where $\Phi$ is the magnetic flux, and hence is not suitable in
  modeling inelastic scattering whereas Buttiker's model preserves the
  Onsager's symmetry relations in the presence of magnetic flux.
  However, the Buttiker model suffers from a shortcoming, in that it
  describes only localized dephasing (at the point contact between
  system and third lead) instead of dephasing that occurs throughout
  the system in realistic situations.

\begin{figure}[h]
\protect\centerline{\epsfxsize=3.0in \epsfbox{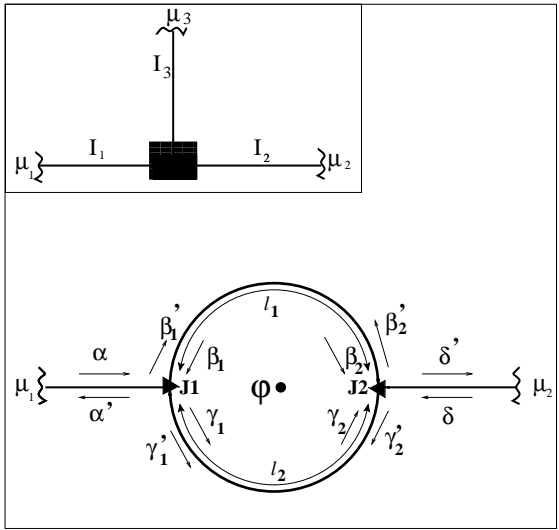}}
\caption{Aharonov - Bohm ring geometry. Inset shows the three probe model.}
\end{figure}
  
  Brouwer and Beenakker\cite{brouwer} have removed the shortcomings of
  both the above stated models by mapping the three probe Buttiker's
  method into a two terminal geometry, this is done by eliminating the
  transmission coefficients which explicitly depend on the third probe
  by means of unitarity of the S-matrix.  They have considered a three
  terminal geometry in which one of the probes is used as a voltage
  probe in absence of magnetic flux (see inset of Fig.~1).  A current
  $I=I_1=-I_2$ flows from source to drain. In this model, a fictitious
  third lead connects the ring to a reservoir at chemical potential
  $\mu_3$ in such a way that no current is drawn $(I_3=0)$.  The
  $3$X$3$ S-matrix of the entire system can be written as-

\[S=\left(\begin{array}{ccc}
r_{11}& t_{12}&t_{13}\\
t_{21}& r_{22}&t_{23}\\
t_{31}& t_{32}&r_{33}
\end{array} \right) \]
Application of the relations\cite{buti,brouwer,datta}-
$I_p=\sum_q G_{pq}[\mu_p-\mu_q], p=1,2,3$ and 
$ G_{pq}=(2e^2/h)T_{pq}$

yields the (dimensionless) two probe conductance $G=\frac{h}{2e^2}\frac{I}{\mu_1-\mu_2}$,

\begin{eqnarray}
G=T_{21}+\frac{T_{23}T_{31}}{T_{31}+T_{32}}
\end{eqnarray}

and on elimination of the transmission coefficients which involve the
voltage probe using the unitarity of the S - Matrix leads to\cite{brouwer}

\begin{eqnarray}
G=T_{21}+\frac{(1-R_{11}-T_{21})(1-R_{22}-T_{21})}{1-R_{11}-T_{21}+1-R_{22}-T_{12}}.
\end{eqnarray}

Now all the above coefficients are built from the 2X2 S matrix-
\[S^{\prime}=\left(\begin{array}{cc}
r_{11}& t_{12}\\
t_{21}& r_{22}\\
\end{array} \right) \]
which represents the S Matrix of the absorbing system.  The first term
in Eq.~2 represents the conductance contribution from the phase
coherent part. The second term accounts for electrons that are
re-injected from the phase breaking reservoir, thereby ensuring
particle conservation in the voltage probe model. For further
calculation one can use the coefficients of the $S^{\prime}$ matrix of
an absorbing system where absorption takes place uniformly.

There are different ways of introducing absorption in the system -(1)
The well known imaginary potential (IP) model and (2)Stochastic
absorption (SA). In the first case Hamiltonian becomes non-hermitian
due to the presence of complex potential term $-i V_i$, which leads to
absorption (non conservation of particle number).  This model suffers
from some spurious features\cite{jayan,rubio}. In the scattering case, in the vicinity
of the absorber, the particle experiences a mismatch in the potential
(being complex) and therefore it tries to avoid this region by
enhanced back reflection. Thus the IP model plays a dual role of an
absorber as well as a reflector, i.e., absorption without reflection
is not possible. Naively one expects the absorption to increase
monotonically as a function of $V_i$. However, the observed behavior
is non-monotonic. At first absorption increases and after exhibiting a
maximum decreases to zero as $V_i\rightarrow\infty$. The absorber in
this limit acts as a perfect reflector. During each scattering event
an electron picks up an additional scattering phase shift due to $V_i$
which along with multiple reflections leads to spurious scatterings
(additional resonances)\cite{jayan,rubio} in the system. As a
result of the afore mentioned limitations there is need of a model
wherein such spurious scatterings are absent and in the limit of large
absorption does not correspond to a perfect reflector. Fortunately
such a model exists, and this is the model of stochastic
absorption\cite{joshi}. This model does not suffer from the
drawbacks of the IP model as we will show towards the end of our work.

The SA model is not new it has earlier been dealt with in the context
of ID localization\cite{joshi}. Stochastic absorption in the ring is
inserted by the factor $e^{-\alpha l_1}$ (or $e^{-\alpha l_2}$) in the
complex free propagator amplitudes, every time we traverse\cite{datta}
the upper (or lower) arms of the ring (see Fig.~1). We have
calculated the relevant transmission and reflection coefficients by
using the S matrix method along with the quantum wave guide theory for a
single channel case. In this model, average absorption per unit length
is given by $2\alpha$.  With this method we show that the calculated
conductance $(G)$ in Eq.~2 is symmetric under the flux reversal as
required. The visibility of the AB oscillations rapidly decay as a
function of $\alpha$, indicating dephasing. Hence forth we refer to
$\alpha$ as an incoherence parameter. Increasing $\alpha$ corresponds
to increasing dephasing processes in the system or increase in
temperature.

In Fig.~1, the length of the upper arm is $l_1$ and that of lower arm
is $l_2$. The total circumference of the loop is $L=l_1+l_2$. The loop
is connected to two current leads.  The couplers (triangles) in
Fig.~1 which connects the leads and the loop are described by a
scattering matrix $S$. The S matrix for the left coupler yields the
amplitudes $O_{1}=(\alpha^\prime,\beta_{1}^\prime,\gamma_{1}^\prime)$
emanating from the coupler in terms of the incident waves
$I_1=(\alpha,\beta_{1},\gamma_{1})$, and for the right coupler yields
the amplitudes
$O_{2}=(\delta^\prime,\beta_{2}^\prime,\gamma_{2}^\prime)$ emanating
from the coupler in terms of the incident waves
$I_2=(\delta,\beta_{2},\gamma_{2})$. The S-matrix for either of the
couplers\cite{butipra} is given by-

\[S=\left(\begin{array}{ccc}
-(a+b)       & \sqrt\epsilon&\sqrt\epsilon\\
\sqrt\epsilon& a            &b            \\
\sqrt\epsilon& b            &a            
\end{array} \right) \]

with $a=\frac{1}{2}(\sqrt{(1-2\epsilon)} -1)$ and
$b=\frac{1}{2}(\sqrt{(1-2\epsilon)} +1)$. $\epsilon$ plays the role of
a coupling parameter. The maximum coupling between reservoir and loop
is $\epsilon=\frac{1}{2}$, and for $\epsilon=0$, the coupler
completely disconnects the loop from the reservoir. As mentioned above
incoherence is taken into account by introducing an attenuation
constant per unit length as mentioned before.

The waves incident into the branches of the loop are related by the S
Matrices for upper branch by-

\[\left(\begin{array}{c}
\beta_1\\
\beta_2\\
\end{array} \right) \ =\left(\begin{array}{cc}
0     & e^{ikl_1} e^{-\alpha l_1} e^\frac{-i \theta l_1}{L}\\
e^{ikl_1} e^{-\alpha l_1} e^\frac{i \theta l_1}{L} & 0 \\
\end{array} \right) \left(\begin{array}{c}
\beta_1^\prime\\
\beta_2^\prime
\end{array} \right)\]  
and  for lower branch-

\[\left(\begin{array}{c}
\gamma_1\\
\gamma_2\\
\end{array} \right) \ =\left(\begin{array}{cc}
0     & e^{ikl_2} e^{-\alpha l_2} e^\frac{i \theta l_2}{L}\\
e^{ikl_2} e^{-\alpha l_2} e^\frac{-i \theta l_2}{L} & 0 \\
\end{array} \right) \left(\begin{array}{c}
\gamma_1^\prime\\
\gamma_2^\prime
\end{array} \right)\]
These S matrices of course are not unitary $
S(\alpha)S(\alpha)^\dagger\neq 1 $ but they obey the relation $
S(\alpha)S(-\alpha)^\dagger= 1 $. The same relation is also obeyed by
the S Matrix of the system in presence of imaginary potential.  Here
$kl_1$ and $kl_2$ are the phase increments of the wave function in
absence of flux.  $\frac{\theta l_1}{L}$ and $\frac{\theta l_2}{L}$
are the phase shifts due to flux in the upper and lower branches.
Clearly, $\frac{\theta l_1}{L}+\frac{\theta
  l_2}{L}=\frac{2\pi\Phi}{\Phi_0} $, where $\Phi$ is the flux piercing
the loop and $\Phi_0$ is the flux quantum$\frac{hc}{e}$. The
transmission and reflection coefficients in Eq.~2 are given as
follows- $T_{21}=|\frac{\delta^\prime}{\alpha}|^2$,
$R_{11}=|\frac{\alpha^\prime}{\alpha}|^2$,
$R_{22}=|\frac{\delta^\prime}{\delta}|^2$,
$T_{12}=|\frac{\alpha^\prime}{\delta}|^2$ wherein
$\delta^\prime,\delta,\alpha^\prime,\alpha$ are as depicted in Fig.~1.

After calculating the required reflection and transmission
coefficients we graphically represent our results in the following
figures.  Throughout the discussion the physical parameters we give
are in dimensionless units.  We see that the coherent
transmission $T_{21}$ is not symmetric
under the flux reversal however proper re-injection of carriers by Eq.~2 for
the total conductance $G$ plotted in Fig.~2 shows that the Onsager's
symmetry relations are restored, i.e., $G$ is symmetric under flux
reversal. In the inset of Fig.~2 we have plotted $G$ versus flux for
the same physical parameters except we have considered the weak
coupling case $(\epsilon=0.10)$. $G$ is symmetric in flux and features
are sharp as expected because of lifetime broadening of the energy
levels is small in this case compared to the case for which
$\epsilon=0.44$.

\begin{figure} [h]
\protect\centerline{\epsfxsize=3.5in\epsfbox{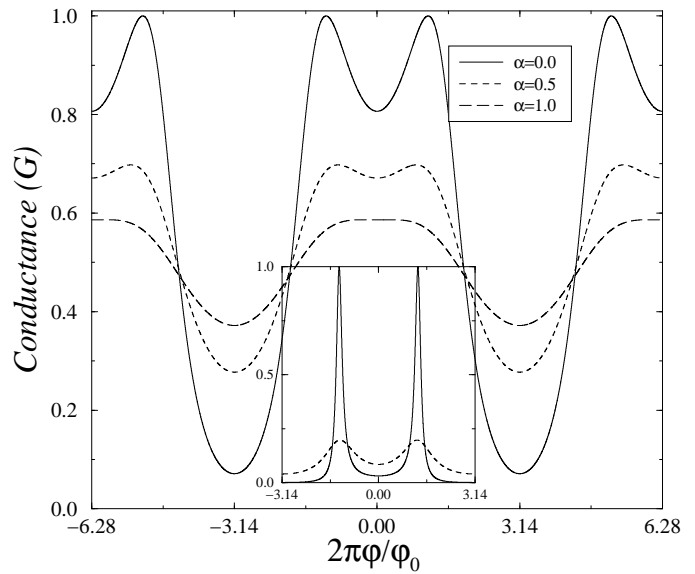}}
\caption{Conductance ($G$)for lengths $l_{1}/l=0.45$, $l_{2}/l=0.55$
  for coupling parameter $\epsilon=0.44$ (waveguide coupling).
  Fermi wave-vector $kl=5.0$. The inset shows the weak coupling case
  $\epsilon=0.10$ for same physical parameters. }
\end{figure}

In Fig.~3 we plot visibility ($V$) as a function of incoherence parameter
$\alpha$. Visibility is of course defined as-
\begin{eqnarray}
V=\frac{G_{max}-G_{min}}{G_{max}+G_{min}}.
\end{eqnarray}

\begin{figure} [h]
\protect\centerline{\epsfxsize=3.5in\epsfbox{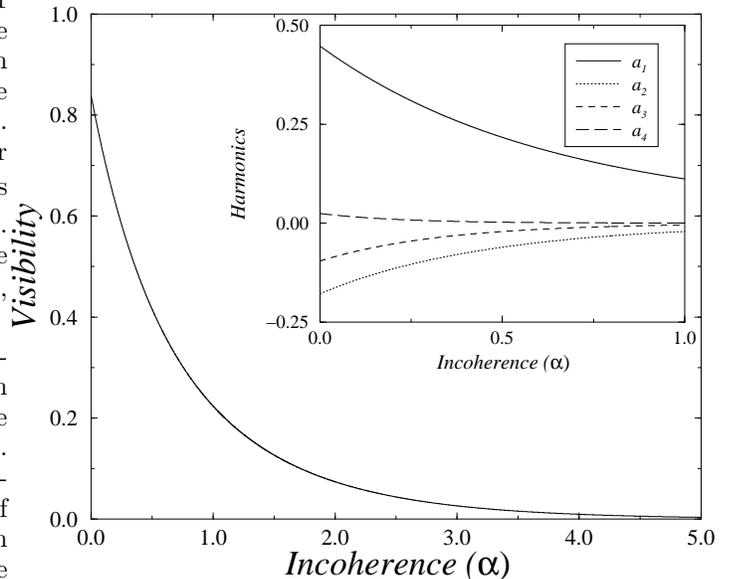}}
\caption{Visibility
 for the same physical parameters as in figure~2, coupling parameter
 $\epsilon=0.44$. In the inset the harmonics have
 been plotted for the same physical and coupling parameters.}
\end{figure}

The plot shows that with increase in the value of the parameter
$\alpha$ the visibility exponentially falls off, reaching a point where it 
becomes zero corresponding to the
disappearance of quantum interference effects. In the inset
of Fig.~3 we have plotted the first few Fourier\cite{xia} harmonics
$a_i$ (wherein $i=1$ to $4$) of $G(\Phi)$ as a function of $\alpha$.
The harmonics exponentially fall off with increasing $\alpha$ with
exponent increasing as we go from $1$ to $4$. The $nth$ order harmonic
corresponds to the contribution from electronic paths which encircle
the flux $n$ times.  The harmonics can sometimes show non monotonic
behavior depending on the physical parameters, however, the visibility
is a monotonic function of the incoherence parameter $\alpha$.

\begin{figure} [h]
\protect\centerline{\epsfxsize=3.5in\epsfbox{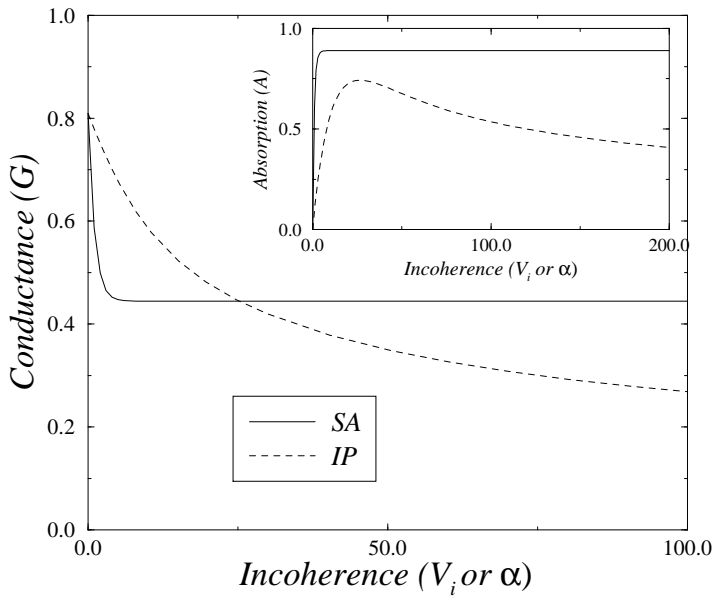}}
\caption{Comparison of two models of dephasing. Herein the
  conductance (absorption in the inset) for the two cases $SA$ and $IP$ have been plotted. The parameters used are same as in Fig.~3.}
\end{figure}

Now for the comparison between the two models, i.e., IP and SA, in
Fig.~4 we plot the total conductance $(G)$ for both the cases. In the
case of imaginary potential we see that the total conductance $(G)$
continuously decreases and as $V_i\rightarrow\infty$, $G\rightarrow 0$
(this limit is not shown in the graph as it goes beyond the scale
used), while in-case of stochastic absorption as
$\alpha\rightarrow\infty$, $G$ goes to a constant value which depends
on the Fermi-energy and other system parameters.  Thus we see that the
modeling of dephasing by SA is indeed justified as we want the AB
oscillations to die out and not that the conductance itself should
vanish, and this is where SA scores over the IP model. In the figure
inset we have depicted the behavior of total absorption in the system
$A$ for both the cases for the same physical parameters. As
$\alpha\rightarrow\infty$ there is a finite absorption in the system
as the electron propagates in the medium in this limit whereas
absorption in the imaginary potential model is non-monotonic and in
the limit $V_i\rightarrow\infty$ absorption vanishes. This is due to
the fact that in the IP model as $V_i\rightarrow\infty$ absorber acts
as a perfect reflector, there is no absorption in the medium as the
particles do not enter the medium (and hence $G=0$) obviously which is
an unrealistic situation for real systems.

In conclusion, we have shown that $G(\Phi)$ is symmetric under flux
reversal in the presence of incoherent scattering represented by the
incoherence parameter $\alpha$. For this we have used the
procedure of Brouwer and Beenakker in the presence of magnetic flux
and simulated absorption (dephasing) using the method of stochastic
absorption. We have used this method of SA to study the behaviors of
the various quantum phenomena, e.g., transport across resonant
tunneling systems, current magnification effect in mesoscopic
rings,\cite{psdeo} delay and sojourn times\cite{anantha} in mesoscopic systems in the presence
of incoherence. These results will be published elsewhere.

\begin{acknowledgements}
One of us (AMJ) thanks Professor Markus Buttiker for his communication and his views on this problem. 
\end{acknowledgements}

\end{document}